\documentstyle[amssymb,aps]{revtex}

\begin{document}

\noindent {\bf A }\noindent {\bf DETAILED\ RE}\noindent \noindent \noindent
\noindent \noindent \noindent {\bf CALCULATION OF THE SPECTRUM OF THE
RADIATION EMITTED\ DURING\ GRAVITATIONAL\ COLLAPSE OF\ A\ SPHERICALLY\
SYMMETRIC\ STAR}

\bigskip

\bigskip \noindent \noindent {\bf A\ Calogeracos}\noindent (*)

\noindent {\it Division of Theoretical Mechanics, Hellenic Air Force Academy
TG1010, Dekelia Air Force Base, Greece}

\medskip

\noindent

\noindent

\noindent

\noindent \noindent July 2003

\bigskip \noindent (*) acal@hol.gr

\medskip

\noindent

\begin{center}
\bigskip
\end{center}

{\bf Abstract}

\bigskip

We address the question of radiation emission from a collapsing star. We
consider the simple model of a spherical star consisting of pressure-free
dust and we derive the emission spectrum via a systematic asymptotic
expansion of the complete Bogolubov amplitude. Inconsistencies in
derivations of the black body spectrum are pointed out.

\section{\protect\bigskip Introduction}

More than a quarter of a century has passed since Hawking's remarkable
suggestion that a star collapsing to a black hole gives rise to radiation
emission at a steady rate characterized by the black body spectrum (Hawking
1975). That particle creation takes place is not in itself surprising. Let
us take the case of a spherical star with radius initially larger than its
Schwarzschild radius that eventually collapses contracting to a point
(according to classical gravitation). If we consider a quantized scalar
photon field the quantum spaces of the $in$ states (before the initiation of
collapse) and of the $out$ states (after collapse is completed) are
certainly different. Hence particle creation clearly takes place and is
determined by the Bogolubov $\alpha (\omega ,\omega ^{\prime })$ and $\beta
(\omega ,\omega ^{\prime })$ amplitudes. The fact that the spectrum is that
of a black body is indeed noticeable, and ties up with the somewhat earlier
results on black hole thermodynamics. From the mathematical point of view
the black body result is due to the special behaviour of the photon modes
near the stellar surface just before the horizon is formed. Hawking in his
derivation makes heavy use of the asymptotic (i.e. near the horizon) form of
the modes. The singular behaviour of the modes in this regime has in turn
given rise to various statements in the literature that are not strictly
correct. For example there have been references to ''parts of $\alpha
(\omega ,\omega ^{\prime })$ and $\beta (\omega ,\omega ^{\prime })$ that
relate to the steady-state regime at late times'' ((DeWitt 1975), p. 327).
However the Bogolubov amplitudes are global constructs and the distinction
between early and late times does not make strict sense. Hawking too talks
about particle production that depends on the details of the collapse and
that such particles ''will disperse'', thus leaving only the thermal part at
late times (Hawking (1975), p. 207). The point of view we are adopting in
this paper is the one dictated by quantum-mechanical orthodoxy, namely that
particle production is a global process ((Hawking 1975) p. 216) and that one
should start with the full standard expression for $\beta (\omega ,\omega
^{\prime })$. Of course at some point in the mathematical handling of the
amplitude $\beta (\omega ,\omega ^{\prime })$ \noindent the special role of
the horizon will show up. The above remarks are certainly not meant to imply
that one cannot calculate {\it local} quantities like $\left\langle T_{\mu
\nu }\right\rangle $; such quantities certainly behave very differently
during the various phases of the collapse.

Shortly after Hawking's work on black hole evaporation two classic papers
(Fulling and Davies 1976) and (Davies and Fulling 1977) were published (the
latter shall be quoted as DF in what follows). The authors demonstrated an
illuminating analogy, physical as well as mathematical, between
gravitational collapse and the seemingly rather different problem of a
perfect mirror starting from rest and accelerating for an infinite time. The
renormalized matrix element of the $T_{uu}$ component of the energy momentum
tensor, defined as 
\[
T_{uu}=\left( \partial _{u}\phi \right) ^{2} 
\]
\noindent is calculated in DF. The result is that {\it asymptotically for }$%
t\rightarrow \infty $%
\begin{equation}
\left\langle T_{uu}\right\rangle \rightarrow \frac{\kappa ^{2}}{48\pi }
\label{bb1a}
\end{equation}
(see e.g. equation (4.16) of (Birrell and Davies 1982) where a comprehensive
review of particle production from both mirrors and black holes is
presented; $\kappa $ is a constant characterizing the trajectory.) Equation (%
\ref{bb1a}) shows that there is a constant energy flux at late times,
analogous to the thermal energy flux found in (Hawking 1975). Davies and
Fulling (1977) also calculated the Bogolubov amplitude following Hawking's
lines and arrived at the black body spectrum relevant to an accelerating
mirror 
\begin{equation}
\left| \beta (\omega ,\omega ^{\prime })\right| ^{2}=\frac{1}{2\pi \omega
^{\prime }\kappa }\frac{1}{e^{2\pi \omega /\kappa }-1}  \label{bb1c}
\end{equation}

\begin{equation}
n(\omega )=\int_{0}^{\infty }d\omega ^{\prime }\left| \beta (\omega ,\omega
^{\prime })\right| ^{2}  \label{bb2}
\end{equation}

\noindent Certain technicalities in the latter calculation have been
elucidated in a previous paper by the author (Calogeracos 2002, hereafter
referred to as I). In the present paper we intend to show that similar
points may be raised in connection to the standard black hole literature
(see (Birrell and Davies 1982), chapter 8 and references therein). There are
steps in the derivation of the black body spectrum that both obscure the
mathematics and create confusion regarding the physics of the problem. Our
findings, summarized in the concluding section, in no way do they diminish
the astounding intuition shown by the early workers on the subject.\noindent

In section 2 we examine the simple model of a collapsing spherical star
consisting of pressure-free dust and we review the fundamental features of
the collapse. The problem is widely treated in the literature and some of
the results are included so that the paper be self-contained. In section 3
we write down the photon modes and the full expression for the Bogolubov
amplitude $\beta (\omega ,\omega ^{\prime })$. In section 4 we show that the
asymptotic behaviour of the latter for large $\omega ^{\prime }$ is 
\begin{equation}
\beta (\omega ,\omega ^{\prime })\approx \left( \omega ^{\prime }\right) ^{-%
\frac{1}{2}}+O\left( \left( \omega ^{\prime }\right) ^{-N}\right) \left(
N>1\right)  \label{bb2a}
\end{equation}

\noindent The $1/\omega ^{\prime }$ in (\ref{bb1c}) leads to a logarithmic
divergence in (\ref{bb2}) and this signals the production of particles at a
finite rate for an infinite time. One often refers to the ultraviolet
divergence by saying that large $\omega ^{\prime }$ frequencies dominate.
Our analysis emphasizes two points, stressed also in I: (a) the thermal
result depends crucially on the behaviour of the photon modes near the
horizon, (b) for a consistent derivation one must consider the {\it whole }%
collapse and {\it not }just the late phase. The truth of statement (a) is
usually taken as common knowledge. However the importance of statement (b)
is often not appreciated. Comparison of our approach with the somewhat more
conventional one is presented in Appendix A.

The ultraviolet divergence previously mentioned is widely taken to signify
that one cannot fully resolve the problem within the context of classical
gravitation. It is well known that the constants we have at our disposal are 
$c=3\times 10^{10}$ cm$\cdot $s$^{-1}$, $\not{h}=1.05\times 10^{-27}$ g$%
\cdot $cm$^{2}\cdot $s$^{-1}$, $G=6.67\times 10^{-8}$ cm$^{3}\cdot $s$%
^{-2}\cdot $g$^{-1\text{ }}$and that we can form three quantities with
dimensions of mass, length, and time respectively: $M=\left( \not%
{h}c/G\right) ^{1/2}=2\times 10^{-5}$ g, $L=\left( \not{h}G/c^{3}\right)
^{1/2}=1.6\times 10^{-33}$ cm, $T=\left( \not{h}G/c^{5}\right)
^{1/2}=5\times 10^{-44}$ s. It is clear that when the radius of the
contracting star becomes of order $L$ then quantum gravitational effects
have to come into play. Such matters are not discussed here. Nor do we touch
upon the profound problem of loss of information during collapse (see e.g.
Preskill 1992, Helfer 2003).

Note: in what follows $\not{h}=G=c=1$.

\section{A collapsing sphere of dust}

\medskip

We consider a collapsing spherical star consisting of pressure-free dust and
follow the approach of (Weinberg 1972), chapter 11, sections 8 and 9. Since
each dust particle falls freely the spacetime geometry inside the star is
most appropriately described in a comoving frame. Let $M$ be the mass of the
star. The metric reads 
\begin{equation}
ds^{2}=dt^{2}-R^{2}(t)\left[ \frac{dr^{2}}{1-kr^{2}}+r^{2}d\theta
^{2}+r^{2}\sin ^{2}\theta d\varphi ^{2}\right]  \label{c1}
\end{equation}

\noindent In the comoving frame each dust particle is labelled by a unique
set of $(r,\theta ,\varphi )$ and the time $t$ stands for the proper time
registered by the particle in question. The radius of the star is specified
in the comoving frame by $r=a$ where $a$ is by definition constant. The
quantities $k$ and $R(t)$ refer to details of the model and their meaning
shall be explained shortly. The collapse is initiated at time $t=0$, and we
normalize $R(t)$ by requiring 
\begin{equation}
R(0)=1  \label{co4}
\end{equation}
Outside the star spacetime is described by the Schwarzschild metric 
\begin{equation}
ds^{2}=\left( 1-\frac{2M}{\bar{r}}\right) d\bar{t}^{2}-\left( 1-\frac{2M}{%
\bar{r}}\right) ^{-1}d\bar{r}^{2}-\bar{r}^{2}d\bar{\theta}^{2}-\bar{r}%
^{2}\sin ^{2}\bar{\theta}d\bar{\varphi}^{2}  \label{c2}
\end{equation}

\noindent The match of the two metrics at the surface implies (Weinberg
1972) 
\begin{equation}
\bar{r}=rR(t),\theta =\bar{\theta},\varphi =\bar{\varphi}  \label{co3}
\end{equation}

\noindent Relation (\ref{co4}) together with the first of (\ref{co3}) imply
that the quantity $a$ stands for the radius of the star in Schwarzschild
spacetime at the moment when collapse starts. Thus the radius of the star in
Schwarzschild coordinates is given by $\bar{R}=aR$. The constant $k$ is
proportional to the star's density and is related to the mass via 
\begin{equation}
2M=ka^{3}  \label{co5}
\end{equation}

\noindent At times we shall consider the case of an initially very large
star $a>>M$, which implies via (\ref{co5}) 
\begin{equation}
ka^{2}<<1  \label{co31}
\end{equation}
The relation between $\bar{t}$ and $t$ is a rather more intricate one and is
treated in detail by Weinberg {\it op. cit. }(see the remark after (\ref{co8}%
) below). It has already been mentioned that since the pressure inside the
star vanishes the motion of the surface corresponds to that of a free
falling particle in the gravitational field of a body of mass $M$. The
proper time $t$ registered by an observer located on the surface of the
collapsing star and the quantity $\bar{R}$ are best given parametrically in
terms of the cycloidal variable $\eta $ (Weinberg 1972): 
\begin{equation}
\bar{R}(\eta )=aR(\eta ),R(\eta )=\frac{1}{2}\left( 1+\cos \eta \right)
\label{co6}
\end{equation}

\begin{equation}
t(\eta )=\frac{1}{2\sqrt{k}}\left( \eta +\sin \eta \right)  \label{co7}
\end{equation}

\noindent where $0\leq \eta \leq \pi $. Collapse starts at $\eta =0$ and is
completed at $\eta =\pi $. Thus according to (\ref{co7}) the total time of
the collapse as measured by an observer comoving with the star's surface is
given by $\pi /\left( 2\sqrt{k}\right) $. The black hole is formed when $%
\bar{R}=2M$. According to (\ref{co6}) this happens at a value $\eta _{0}$ of
the parameter given by 
\begin{equation}
\frac{\eta _{0}}{2}=\cos ^{-1}a\sqrt{k}  \label{co9}
\end{equation}
The Schwarzschild time $\bar{t}$ for a particle on the stellar surface is
given in terms of $\eta $ by the expression (see (Chandrasekhar 1983)
chapter 3, section 19) : 
\begin{equation}
\bar{t}=2a\sqrt{1-ka^{2}}\left[ \frac{1}{2}\left( \eta +\sin \eta \right)
+ka^{2}\eta \right] +2M\ln \left[ \frac{\tan \frac{\eta _{0}}{2}+\tan \frac{%
\eta }{2}}{\tan \frac{\eta _{0}}{2}-\tan \frac{\eta }{2}}\right]  \label{co8}
\end{equation}

\noindent Relations (\ref{co7}) and (\ref{co8}) parametrically connect the
time $t$ in the comoving frame to the Schwarzschild time $\bar{t}$ on the
stellar surface. Note that as $\eta \rightarrow \eta _{0}$ from the left
(i.e. before the black hole is formed) $t$ diverges. This reflects the
well-known fact that collapse appears to the Schwarzschild observer to take
an infinitely long time.

Radial null geodesics in Schwarzschild spacetime satisfy (see (\ref{c2})) 
\begin{equation}
dt^{2}=\frac{1}{\left( 1-\frac{2M}{\bar{r}}\right) ^{2}}d\bar{r}^{2}
\label{c10}
\end{equation}

\noindent This leads to the definition of the Regge-Wheeler radial
coordinate $r^{*}$%
\begin{equation}
r^{*}=\bar{r}+2M\ln \left( \frac{\bar{r}}{2M}-1\right)  \label{co11}
\end{equation}

\noindent so that (\ref{c10}) may be written as $d\bar{t}^{2}-dr^{*2}=0$,
and to the introduction of the incoming and outgoing Eddington-Finkelstein
coordinates 
\begin{equation}
v=\bar{t}+r^{*},u=\bar{t}-r^{*}  \label{co12}
\end{equation}

\noindent The quantities $v$ and $u$ label photon trajectories in
Schwarzschild spacetime.

In order not to confuse the reader with the proliferation of radial
variables let us clarify: $\bar{R}$ stands for the stellar radius in
Schwarzschild coordinates, $R^{*}$ is the corresponding Regge-Wheeler
coordinate, $R$ is the auxiliary function given by the second of (\ref{co6})
and appearing in (\ref{c1}), $\bar{r}$ is the radial Schwarzschild
coordinate of an arbitrary point outside the star, $r^{*}$ the corresponding
Regge-Wheeler coordinate, and $r$ the radial coordinate inside the star in
the comoving frame.

A property that one must establish is that as the star approaches collapse 
\begin{equation}
\frac{\bar{R}}{2M}-1\simeq Ae^{-\bar{t}/2M}  \label{co14}
\end{equation}

\noindent where $A$ is a calculable constant. This expression is of course
widely known. It agrees with the remark following (\ref{co8}) that as far as
the Schwarzschild observer is concerned the collapse takes an infinitely
long time. A similar exponential appears in the calculation of the red shift
as observed by the Schwarzschild observer; see (Weinberg 1972), p. 348 (the
asymptotic form of the mirror trajectory given by DF, equation (2.3), has a
form identical to that of (\ref{co14})). To establish the validity of (\ref
{co14}) and also calculate $A$ one takes logarithms of both sides and uses (%
\ref{co6}), (\ref{co8}). The logarithmic divergences cancel and $A$ is
expressed in terms of $\eta _{0}$ (see Appendix A). The quantity $A$ depends
on the parameters $a$ and $M$ of the collapse and does not turn up in the
expression for the spectrum. Note that relations (\ref{co12}) define
parametrically (via $\eta $) a function 
\begin{equation}
u=f_{st}(v)  \label{co14b}
\end{equation}

\noindent and its inverse 
\begin{equation}
v=p_{st}(u)  \label{co14c}
\end{equation}

\noindent which give the trajectory of the stellar surface in terms of $u,v$
coordinates.

In the limit $\eta \rightarrow \eta _{0}$ both $R^{*}$ and $\bar{t}$ are
singular (as is obvious from (\ref{co8}) and (\ref{co11})), however their
combination in the Eddington-Finkelstein coordinate $v$ is analytic. The
cancellation of the logarithmic singularities is hardly surprising since $%
v_{0}$ must have a well-defined value labelling a null line; for details see
Appendix A. For given values of $M$ and $a$ the value of $\eta _{0}$ is
readily calculable via (\ref{co9}) and then so is $v_{0}$ after some algebra
(setting $\eta =\eta _{0}$ in (\ref{a3})).Thus $v_{0}-v$ admits a Taylor
expansion near $\eta _{0}$ and of course so does $2M-\bar{R}$ which is
analytic for all $0<\eta <\pi $. Writing 
\begin{equation}
2M-\bar{R}\simeq C_{1}\left( \eta _{0}-\eta \right) ,v_{0}-v\simeq
C_{2}\left( \eta _{0}-\eta \right)  \label{co33}
\end{equation}
we can read the positive constants $C_{1},C_{2}$ off (\ref{a5}), (\ref{a4}).
Then 
\begin{equation}
\bar{R}-2M\simeq C\left( v_{0}-v\right) ,C=C_{1}/C_{2}  \label{co34}
\end{equation}
Combining the above with (\ref{co14}) we get 
\begin{equation}
v\simeq v_{0}-\frac{2MA}{C}e^{-\bar{t}/2M}  \label{co35}
\end{equation}
For $\eta \rightarrow \eta _{0}$ we can combine the two relations (\ref{co12}%
) and write 
\begin{equation}
\bar{t}\simeq \frac{u+v_{0}}{2}  \label{co18}
\end{equation}
Then (\ref{co35}) reads 
\begin{equation}
v\simeq v_{0}-Be^{-u/4M},B\equiv \frac{2MA}{C}  \label{co36}
\end{equation}

\noindent This defines $p_{st}(u)$ in (\ref{co14c}).

In what follows we shall need the metric (\ref{c2}) expressed in terms of
Kruskal coordinates; see e.g. (Townsend 1997), (Misner, Thorne, Wheeler
1973). We introduce 
\begin{equation}
{\it U}=-Ee^{-u/4M},{\it V}=\frac{e^{v/4M}}{E}  \label{co50}
\end{equation}

\noindent and we transform the metric to 
\begin{equation}
ds^{2}=\frac{32M^{3}}{\bar{r}}e^{-\bar{r}/2M}d{\it U}d{\it V}-\bar{r}%
^{2}d\Omega  \label{co51}
\end{equation}

\noindent where $r$ is given in terms of ${\it U,V}$ implicitly via 
\begin{equation}
{\it UV}=\frac{\bar{r}-2M}{2M}e^{\bar{r}/2M}  \label{co52}
\end{equation}

\noindent At early times spacetime is taken to be essentially flat (cf
remark preceding (\ref{co21})) and the metric is simply expressed in terms
of advanced and retarded Eddington-Finkelstein coordinates 
\begin{equation}
ds^{2}=dudv-\bar{r}^{2}d\Omega  \label{co53}
\end{equation}

We now turn to ray-tracing and consider light rays obeying the condition
that they be reflected at the centre of the star $r=0$. The rationale for
the boundary condition at $r=0$ is reviewed in the next section. An incident
ray corresponding to a certain $v$ at early times upon reflection becomes an
outgoing ray corresponding to a certain $u$ at late times, thus defining a
function $u=f(v)$. The inverse function is given by $v=p(u)$. Note that the
function $u=f(v)$ is by construction quite distinct from the function $%
u=f_{st}(v)$ (\ref{co14b}) (the former involves reflection of the ray at the
centre of the star whereas the latter at the surface). The importance of ray
tracing lies in the fact that once we determine the function $f(v)$ (or its
inverse) we can immediately construct the photon modes via (\ref{e3}), (\ref
{e5}). Note also that these expressions involve the Eddington-Finkelstein
coordinates rather than the Kruskal ones. It is convenient to exhibit
collapse via a Kruskal diagram, where radial null lines are at $45^{0}$ on
the plane of the paper. The line $l(1)$ represents a ray that crosses the
star's surface at point B, is reflected at the centre of the star at
spacetime point C, and after reflection crosses the star's surface at A
(reflected ray labelled by $l^{\prime }(1)$). By definition ray $l(1)$ is
the last incoming ray that manages to escape before the black hole is
formed, and corresponds to an incoming Eddington-Finkelstein coordinate $%
v_{H}$. At A the star's surface crosses the Schwarzschild radius, and the
coordinates of A correspond to the value $\eta _{0}$ of the cycloidal
parameter $\eta $ in (\ref{co6}), (\ref{co7}). The extension of the line CA
is the future horizon and corresponds to the value ${\it U}=0$ of the
Kruskal coordinate. Let $l^{\prime }(i)$ be an outgoing ray slightly
preceding $l(1)$, corresponding to an incoming ray $l(i)$. We also trace the
incoming light ray $m(1)$ which becomes $l(1)$ upon reflection on the star's
surface at A without entering the star (this is an auxiliary ray and the
reflection mechanism is an entirely hypothetical one, not connected to
partial reflection that may take place on the star's surface). Similarly let 
$m(i)$ be an incoming ray slightly preceding $m(1)$. The problem is to
relate the $u$ of the outgoing $l^{\prime }(i)$ to the $v$ of the incoming $%
l(i)$, thus determining the function $u=f(v).$ We review the argument in
(Hawking 1975); see also (Townsend 1997), p. 125. Recall ((Townsend 1997),
p. 29 for proof) that ${\it U}$ is an affine parameter for the null geodesic 
${\it U}=${\it constant}. (This means that ${\bf L\cdot D}L^{\mu }=0$ rather
than ${\bf L\cdot D}L^{\mu }$ ${\bf \varpropto }L^{\mu }$.) Instead of ${\it %
U}=-e^{-u/4M}$ one may also use any parameter $\lambda =-Ee^{-u/4M}$, $E$
being some constant; we exploit this freedom and will comment on the value
of $E$ later on. Let ${\bf T}$ be a null vector on ${\it U}=0$ parallel to
the horizon and pointing along the radial spatial direction, and let ${\bf N}
$ be a null vector pointing to the future and along the radial spatial
direction and satisfying ${\bf N\cdot T}=-1$. The above requirements are
consistent (see (Townsend 1997), p. 109) if ${\bf N}$ is parallely
transported along the geodesic defined by ${\bf T}$. Consider a neighbouring
null geodesic labelled by ${\it U}=-\varepsilon $ where is small. This
corresponds to constant $u$ according to (\ref{co50}): 
\begin{equation}
\varepsilon =Ee^{-u/4M}\Rightarrow \ln \varepsilon =\ln E-\frac{u}{4M}
\label{co54}
\end{equation}

\noindent The two geodesics are thus connected by the displacement vector $%
-\varepsilon {\bf N}$. We parallely transport the pair ${\bf N}$, ${\bf T}$
to the point where the ${\it U}=0$ geodesic cuts the trajectory of centre of
the star. We then reflect ${\bf N}$ and ${\bf T}$ as in figure 2 and
parallely transport the new pair to past infinity where the stellar radius
satisfies (\ref{co21}). Then spacetime becomes essentially flat, the metric
is given by (\ref{co53}) and the affine parameter is given by $v$ for a
general geodesic $l(i)$ (figures 1, 2) and by $v_{H}$ for the special
geodesic $l(1)$. We thus have 
\begin{equation}
\varepsilon =v_{H}-v  \label{co55}
\end{equation}

\noindent Comparing (\ref{co54}) and (\ref{co55}) we obtain 
\[
v_{H}-v=Ee^{-u/4M} 
\]

\noindent or 
\begin{equation}
v\simeq v_{H}-Ee^{-u/4M}  \label{co37}
\end{equation}

\noindent (the $\simeq $ symbol above refers to the fact that this
approximation is valid near the horizon). Note that the preceding argument
involving ${\bf N}$, ${\bf T}$ can be equally well applied to the incoming
rays $m(i),$ $m(1)$ which upon reflection on the surface of the star
coincide with $l^{\prime }(i),l^{\prime }(1)$. In particular the affine
distance between $l(1)$ and $l(i)$ equals the distance between $m(1)$ and $%
m(i)$. This implies that the quantity $E$ appearing in (\ref{co37}) is equal
to the $B$ appearing in (\ref{co36}). (The identification of $E$ with $B$
does not affect the derivation of the black body spectrum and we shall keep $%
E$ in notation.) Note also that the quantity

\begin{equation}
d\equiv v_{0}-v_{H}  \label{co13}
\end{equation}

\noindent is calculable; see appendix A, the remark following (\ref{co56}).
Inverting (\ref{co37}) we get 
\begin{equation}
u=f(v)\simeq -\frac{1}{4M}\ln \left( \frac{v_{H}-v}{E}\right)  \label{co38}
\end{equation}

We now turn to the early stage of the collapse (figure 3). We assume that
the collapse is initiated at $\bar{t}=0$ (and that for $\bar{t}<0$ the star
is held stable by some external means). Let us assume large initial stellar
radius 
\begin{equation}
a>>M  \label{co21}
\end{equation}
\noindent Expanding (\ref{co6}), (\ref{co7}) for early times, i.e. small $%
\eta $, we obtain 
\[
\bar{t}\simeq \frac{\eta }{\sqrt{k}},R\simeq 1-\frac{\eta ^{2}}{4} 
\]

\noindent Hence the radius as a function of time at the early stage is given
by 
\begin{equation}
\bar{R}=a\left( 1-\frac{k\bar{t}^{2}}{4}\right)  \label{co20}
\end{equation}

\noindent Recalling the definition (\ref{co5}) we observe that the star's
surface initially contracts according to non-relativistic kinematics with
acceleration $M/a^{2}$ (in accordance with Newton's law of gravity). Given
the assumption (\ref{co21}) the initial gravitational field is sufficiently
weak so the light rays in the Schwarzschild geometry are straight lines. Let
us consider a $\bar{t}-\bar{R}$ diagram ( $\bar{t}$, $\bar{R}$ being
Schwarzschild coordinates; at early times when (\ref{co31}) is satisfied
there is little difference between radial and Regge-Wheeler coordinates) and
take an incoming light ray $q$ which has $v=0$. The reflected ray $q^{\prime
}$ has $u=0$ and takes time $a$ to travel from the centre of the star to the
surface. During this time the gravitational field is still weak. This may be
seen by setting $t=a$ in (\ref{co20}); then (\ref{co31}) implies that by the
time the ray emerges from the star's surface $\bar{R}$ still is almost equal
to $a$. To summarize the function $u=f(v)$ has the properties 
\begin{equation}
f(v)=v=0,initially  \label{co23}
\end{equation}
\begin{equation}
f(v)\simeq -4M\ln \left( \frac{v_{H}-v}{E}\right) ,v\rightarrow v_{H}
\label{co41}
\end{equation}

\section{Construction of the {\it in} and {\it out} states}

We follow (Birrell and Davies 1982), (Brout et al. 1995). A scalar mode
corresponding to angular momentum $l$ and satisfying the Klein-Gordon
equation is written in the form 
\begin{equation}
\psi _{l}=\left( \sqrt{4\pi }r\right) ^{-1}\phi _{l}(r)Y_{l}^{m}(\theta
,\varphi )  \label{co25}
\end{equation}

\noindent where 
\begin{equation}
\left( \frac{\partial ^{2}}{\partial \bar{t}^{2}}-\frac{\partial ^{2}}{%
\partial r^{*2}}-V_{l}(r)\right) \phi _{l}=0  \label{co26}
\end{equation}

\noindent with 
\begin{equation}
V_{l}(r)=\left( 1-\frac{2M}{r}\right) \left( \frac{2M}{r^{3}}+\frac{l(l+1)}{%
r^{2}}\right)  \label{co27}
\end{equation}

\noindent As far as conceptual purposes and technical details are concerned
it suffices to restrict ourselves to $s$ waves (following the references in
the beginning of this section). Note that according to (\ref{co27}) there is
a centrifugal barrier even for $s$ waves which does not affect either the
high frequency modes or the modes that are present at the early stages of
the collapse (when the gravitational field is weak). Again following the
standard treatments of the problem we are going to neglect the barrier. For
a discussion of its effect on the modes see (DeWitt 1975), section 5.2. Thus
(\ref{co26}) becomes 
\begin{equation}
\left( \frac{\partial ^{2}}{\partial \bar{t}^{2}}-\frac{\partial ^{2}}{%
\partial r^{*2}}\right) \phi =0  \label{co29}
\end{equation}
\noindent (in what follows the index $0$ in $\phi $ is suppressed). Since $%
\psi _{0}(r=0)$ must be finite it follows from (\ref{co25}) that 
\begin{equation}
\phi (r=0)=0  \label{co28}
\end{equation}

\noindent Relations (\ref{co29}) and (\ref{co28}) define the problem.

\noindent

Relation (\ref{co29}) can be written in the form 
\begin{equation}
{\displaystyle {\partial ^{2}\phi  \over \partial u\partial v}}%
=0  \label{e1}
\end{equation}

\noindent Hence any function that depends only on $u$ or $v$ (or the sum of
two such functions) is a solution of (\ref{e1}). We can now make contact
with the accelerating mirror problem (DF,I). One set of modes satisfying (%
\ref{e1}) and the boundary condition (\ref{co28}) is given by 
\begin{equation}
\varphi _{\omega }(u,v)=%
{\displaystyle {i \over 2\sqrt{\pi \omega }}}%
\left( \exp (-i\omega v)-\exp \left( -i\omega p(u)\right) \right)  \label{e3}
\end{equation}

\noindent Another set of modes satisfying the boundary condition is
immediately obtained from (\ref{e3}) 
\begin{equation}
\bar{\varphi}_{\omega }(u,v)=%
{\displaystyle {i \over 2\sqrt{\pi \omega }}}%
\left( \exp \left( -i\omega f(v)\right) -\exp \left( -i\omega u\right)
\right)  \label{e5}
\end{equation}

\noindent

The modes $\varphi _{\omega }(u,v)$ of (\ref{e3}) describe sinusoidal waves
incident from $r^{*}=\infty $ as it is clear from the sign of the
exponential in the first term; the second term represents the outgoing part
which has a rather complicated behaviour depending on the motion of the
mirror. These modes constitute the {\it in }space and should obviously be
absent before collapse, 
\begin{equation}
a(\omega ^{\prime })\left| 0in\right\rangle =0  \label{bb4}
\end{equation}
Similarly the modes $\bar{\varphi}_{\omega }(u,v)$ describe sinusoidal
outgoing waves travelling towards $r^{*}=\infty $ (emitted by the star) as
can be seen from the exponential of the second term. Correspondingly the
first term is complicated.

Recall that apart from the modes $\bar{\varphi}_{\omega }(u,v)$ we should
include modes $\bar{q}_{\omega }(u,v)$ that contain no outgoing component;
instead they are confined inside the black hole (the letter $q$ complies
with (Hawking 1975, 1976)). The modes $\bar{q}_{\omega }(u,v)$ are
undetectable by an outside observer, but they are needed to make the set $%
\bar{\varphi}_{\omega },\bar{q}_{\omega }$ complete. Correlations between
the two sets is part of the information problem alluded to in the
Introduction.

The $\bar{\varphi}_{\omega }(u,v)$ modes should be absent from the {\it out }%
vacuum 
\begin{equation}
\bar{a}(\omega )\left| 0out\right\rangle =0  \label{bb6}
\end{equation}
The state \noindent $\left| 0out\right\rangle $ corresponds to the state
where there are no outgoing particles detectable by an outside observer. The
two representations are connected by the Bogolubov transformation 
\begin{equation}
\bar{a}\left( \omega \right) =\int_{0}^{\infty }d\omega \left( \alpha
(\omega ,\omega ^{\prime })a(\omega ^{\prime })+\beta ^{*}(\omega ,\omega
^{\prime })a^{\dagger }(\omega ^{\prime })\right)  \label{e011}
\end{equation}

\noindent Using (\ref{e011}) and its hermitean conjugate we may immediately
verify that the expectation value of the number of excitations of the mode $%
\varphi _{\omega }$ in the $\left| 0in\right\rangle $ vacuum 
\begin{equation}
\left\langle 0in\right| N\left( \omega \right) \left| 0in\right\rangle
=\int_{0}^{\infty }d\omega ^{\prime }\left| \beta (\omega ,\omega ^{\prime
})\right| ^{2}  \label{e0011}
\end{equation}

\noindent The matrix element $\beta (\omega ,\omega ^{\prime })$ is given by
(21) of I (we use $z$ for either $r$ or $r^{*}$ as the case may be to ease
up on notation and make contact with I) 
\begin{equation}
\beta (\omega ,\omega ^{\prime })=-i\int dz\varphi _{\omega ^{\prime }}(z,0)%
{\displaystyle {\partial  \over \partial t}}%
\bar{\varphi}_{\omega }(z,0)+i\int dz\left( 
{\displaystyle {\partial  \over \partial t}}%
\varphi _{\omega ^{\prime }}(z,0)\right) \bar{\varphi}_{\omega }(z,0)
\label{e11}
\end{equation}

\noindent The integration in (\ref{e11}) can be over any spacelike
hypersurface. Since collapse starts at $t=0$ the choice $t=0$ for the
hypersurface is convenient. The $in$ modes evaluated at $t=0$ are given by
the simple expression (\ref{e3}) (i.e. $p(u)=u)$%
\begin{equation}
\varphi _{\omega }(u,v)=%
{\displaystyle {i \over 2\sqrt{\pi \omega }}}%
\left( \exp (-i\omega z)-\exp \left( i\omega z\right) \right) \theta \left(
z\right)  \label{e99}
\end{equation}
where the presence of $\theta \left( z\right) $ emphasizes the fact that $z$
is a radial coordinate. The $\bar{\varphi}$ modes are given by (\ref{e5})
with $f$ depending on the history of the collapse. Relation (\ref{e11}) is
rewritten in the form (the endpoints of integration shall be stated
presently) 
\begin{eqnarray}
\beta (\omega ,\omega ^{\prime }) &=&-i\int dz%
{\displaystyle {i \over 2\sqrt{\pi \omega ^{\prime }}}}%
\left\{ e^{-i\omega ^{\prime }z}-e^{i\omega ^{\prime }z}\right\} \theta
\left( z\right) \frac{\omega }{2\sqrt{\pi \omega }}\left\{ f^{\prime
}e^{-i\omega f}-e^{i\omega z}\right\} +  \label{e100} \\
&&+i\int dz%
{\displaystyle {\omega ^{\prime } \over 2\sqrt{\pi \omega ^{\prime }}}}%
\left\{ e^{-i\omega ^{\prime }z}-e^{i\omega ^{\prime }z}\right\} \theta
\left( z\right) \frac{i}{2\sqrt{\pi \omega }}\left\{ e^{-i\omega
f}-e^{i\omega z}\right\}  \nonumber
\end{eqnarray}

\noindent The above expression may be rearranged in the form 
\begin{equation}
\beta (\omega ,\omega ^{\prime })=%
{\displaystyle {1 \over 4\pi \sqrt{\omega \omega ^{\prime }}}}%
\int_{0}^{v_{H}}dz\left\{ e^{i\omega ^{\prime }z}-e^{-i\omega ^{\prime
}z}\right\} \theta \left( z\right) \left\{ \omega ^{\prime }e^{-i\omega
f}-\omega f^{\prime }e^{-i\omega f}\right\} + 
{\displaystyle {\left( \omega -\omega ^{\prime }\right)  \over 4\pi \sqrt{\omega \omega ^{\prime }}}}%
\int_{0}^{\infty }dz\left\{ e^{i\omega ^{\prime }z}-e^{-i\omega ^{\prime
}z}\right\} \theta \left( z\right) e^{i\omega z}  \label{bb9}
\end{equation}

\noindent where the limits of integration are displayed (we took inro
account the fact that the argument of $f$ runs up to $v_{H}$). The rays with 
$v<0$ do not affect the amplitude due to the presence of the $\theta $
function. The first and second integral in the above relation will be
denoted by $\beta _{1}(\omega ,\omega ^{\prime })$ and $\beta _{2}(\omega
,\omega ^{\prime })$ respectively. Note that the second integration is of
kinematic origin and totally independent of the collapse. We also quote the
expression for the $\alpha (\omega ,\omega ^{\prime })$ amplitude 
\begin{equation}
\alpha (\omega ,\omega ^{\prime })=-i\int_{0}^{\infty }dz\varphi _{\omega
^{\prime }}(z,0)%
{\displaystyle {\partial  \over \partial t}}%
\bar{\varphi}_{\omega }^{*}(z,0)+i\int_{0}^{\infty }dz\left( 
{\displaystyle {\partial  \over \partial t}}%
\varphi _{\omega ^{\prime }}(z,0)\right) \bar{\varphi}_{\omega }^{*}(z,0)
\label{91}
\end{equation}

\noindent Observe the unitarity condition 
\begin{equation}
\int_{0}^{\infty }d\widetilde{\omega }\left( \alpha \left( \omega _{1},%
\widetilde{\omega }\right) \alpha ^{*}\left( \omega _{2},\widetilde{\omega }%
\right) -\beta \left( \omega _{1},\widetilde{\omega }\right) \beta
^{*}\left( \omega _{2},\widetilde{\omega }\right) \right) =\delta \left(
\omega _{1}-\omega _{2}\right)  \label{92}
\end{equation}

\noindent which is a consequence of the fact that the set of {\it in }states
is complete.

Recall from I that we can introduce quantities $A(\omega ,\omega ^{\prime
}),B(\omega ,\omega ^{\prime })$ that are analytic functions of the
frequencies via 
\begin{equation}
\alpha (\omega ,\omega ^{\prime })=\frac{A(\omega ,\omega ^{\prime })}{\sqrt{%
\omega \omega ^{\prime }}},\beta (\omega ,\omega ^{\prime })=\frac{B(\omega
,\omega ^{\prime })}{\sqrt{\omega \omega ^{\prime }}}  \label{e106}
\end{equation}

\noindent The quantity $B(\omega ,\omega ^{\prime })$ is read off (\ref{e100}%
) (and $A(\omega ,\omega ^{\prime })$ from the corresponding expression for $%
\alpha (\omega ,\omega ^{\prime })$). From the definitions of the Bogolubov
coefficients, the explicit form (\ref{e100}) of the overlap integral and
expressions (\ref{e3}) and (\ref{e5}) for the field modes one can deduce
that 
\begin{equation}
B^{*}(\omega ,\omega ^{\prime })=A(-\omega ,\omega ^{\prime }),A^{*}(\omega
,\omega ^{\prime })=B(-\omega ,\omega ^{\prime })  \label{e107}
\end{equation}

\noindent The above relations allow the calculation of $\alpha (\omega
,\omega ^{\prime })$ once $\beta (\omega ,\omega ^{\prime })$ is determined.

\section{Calculation of the Bogolubov amplitudes}

The strategy we adopt in handling (\ref{bb9}) is as follows. The first
integral will be evaluated via an asymptotic expansion in negative powers of 
$\omega ^{\prime }$, which will in fact show that the $\omega ^{\prime }$
integration in (\ref{bb2}) is logarithmically divergent. We start with the
second integral in (\ref{bb9}): 
\begin{equation}
\beta _{2}(\omega ,\omega ^{\prime })= 
{\displaystyle {\left( \omega -\omega ^{\prime }\right)  \over 4\pi \sqrt{\omega \omega ^{\prime }}}}%
\left\{ \int_{0}^{\infty }dze^{i\left( \omega +\omega ^{\prime }\right)
z}-\int_{0}^{\infty }dze^{i\left( \omega -\omega ^{\prime }\right) z}\right\}
\label{co80}
\end{equation}
\noindent The integrals in (\ref{co80}) are readily evaluated in terms of
the function $\zeta $ and its complex conjugate $\zeta ^{*}$(see e.g.
Heitler (1954), pages 66-71): 
\begin{equation}
\zeta (x)\equiv -i\int_{0}^{\infty }e^{i\kappa x}d\kappa =P\frac{1}{x}-i\pi
\delta (x)  \label{delta}
\end{equation}

\noindent Since we are interested in the asymptotic limit $\omega ^{\prime
}\rightarrow \infty $ the argument of the $\delta $ function in (\ref{delta}%
) never vanishes so it is only the first term in (\ref{delta}) that is
operative as far as the calculation of the $\beta _{2}(\omega ,\omega
^{\prime })$ goes. (The $\delta $ proportional term is relevant in the
calculation of the $\alpha (\omega ,\omega ^{\prime })$ amplitude via
relations (\ref{e106}), (\ref{e107}).) Thus asymptotically in the said limit 
\begin{equation}
\beta _{2}(\omega ,\omega ^{\prime })\simeq \frac{1}{2\pi i\sqrt{\omega
\omega ^{\prime }}}  \label{bb10}
\end{equation}

We turn to the first integral $\beta _{1}(\omega ,\omega ^{\prime })$ in (%
\ref{bb9}). Rather than splitting it to four integrals we perform an
integration by parts to get (\ref{bb9}) in the form 
\begin{equation}
\beta _{1}(\omega ,\omega ^{\prime })=-\frac{1}{2\pi }\sqrt{\frac{\omega
^{\prime }}{\omega }}\int_{0}^{v_{H}}dze^{-i\omega f(z)-i\omega ^{\prime }z}+%
\frac{1}{2\pi }\frac{1}{\sqrt{\omega \omega ^{\prime }}}\sin \left( \omega
^{\prime }v_{H}\right) e^{-i\omega f\left( v_{H}\right) }  \label{bb22}
\end{equation}
This is the same integration by parts that was used to obtain (35) of I and
also the one that is used in DF to go from their (2.10a) to (2.10b). The
lower limit contribution of the integration by parts vanishes (being
proportional to $\sin \omega ^{\prime }z$), quite irrespectively of the
value of $f(0)$. The second term in (\ref{bb22}) (originating from the upper
limit) oscillates rapidly since the exponent tends to infinity. Thus the
term tends distributionally to zero it may be neglected as in DF (also it is
one power of $\omega ^{\prime }$ down compared to the first term). So
asymptotically in $\omega ^{\prime }$ we are entitled to write 
\begin{equation}
\beta _{1}(\omega ,\omega ^{\prime })\simeq -\frac{1}{2\pi }\sqrt{\frac{%
\omega ^{\prime }}{\omega }}\int_{0}^{v_{H}}dze^{-i\omega f(z)-i\omega
^{\prime }z}  \label{bb23}
\end{equation}
To bring the singularity in the integral to zero we make the change of
variable

\begin{equation}
z=v_{H}-x  \label{e22a}
\end{equation}

\noindent and rewrite $\beta _{1}(\omega ,\omega ^{\prime })$ in the form 
\begin{equation}
\beta _{1}(\omega ,\omega ^{\prime })=-\frac{e^{-i\omega v_{H}}}{2\pi }\sqrt{%
\frac{\omega ^{\prime }}{\omega }}\int_{0}^{v_{H}}dxe^{-i\omega
g(x)}e^{i\omega ^{\prime }x}  \label{bb24}
\end{equation}

\noindent where the function \noindent \noindent $g(x)\equiv f\left(
v_{H}-z\right) $ is defined in the range $0<x<v_{H}$ and has the properties
that follow from (\ref{co23}), (\ref{co41}): 
\begin{equation}
g(x)\simeq -4M\ln \left( \frac{x}{E}\right) ,x\rightarrow 0  \label{co42}
\end{equation}
\begin{equation}
g(v_{H})=f(0)  \label{co43}
\end{equation}
We isolate the integral

\begin{equation}
I\equiv \int_{0}^{v_{H}}dxe^{-i\omega g(x)}e^{i\omega ^{\prime }x}
\label{e24b}
\end{equation}

\noindent

To obtain the asymptotic behaviour of (\ref{e24b}) for $\omega ^{\prime }$
large we adopt the standard technique of deforming the integration path to a
contour in the complex plane; see (Bender and Orszag 1978), chapter 6;
(Ablowitz and Fokas 1997), chapter 6; (Morse and Feshbach 1953), p. 610
where a very similar contour is used in the study of the asymptotic
expansion of the confluent hypergeometric. The deformed contour runs from 0
up the imaginary axis till $iT$ (we eventually take $T\rightarrow \infty $),
then parallel to the real axis from $iT$ to $iT+v_{H}$, and then down again
parallel to the imaginary axis from $iT+v_{H}$ to $v_{H}$. The contribution
of the segment parallel to the real axis vanishes exponentially in the limit 
$T\rightarrow \infty $. We thus get 
\begin{equation}
I=i\int_{0}^{\infty }dse^{-\omega ^{\prime }s}e^{-i\omega
g(is)}-i\int_{0}^{\infty }dse^{i\omega ^{\prime }\left( v_{H}+is\right)
}e^{-i\omega g(v_{H}+is)}  \label{co44}
\end{equation}
In both integrations the dominant contribution comes from the region where $%
s\simeq 0$. The second integral (including the minus sign in front) takes
the form using (\ref{co43}) 
\begin{equation}
-ie^{i\omega ^{\prime }v_{H}}e^{-i\omega f(0)}\int_{0}^{\infty }dse^{-\omega
^{\prime }s}=-i\frac{e^{i\omega ^{\prime }v_{H}}e^{-i\omega f(0)}}{\omega
^{\prime }}  \label{co45}
\end{equation}

\noindent In the first integral we use the asymptotic form (\ref{co42})
valid for small $x$ and write 
\[
\exp \left( -i\omega g\left( is\right) \right) =\exp \left( i4M\omega \ln
\left( \frac{is}{E}\right) \right) =\left( \frac{is}{E}\right) ^{i4M\omega
}=\exp \left( -\frac{\pi }{2}4M\omega \right) \left( \frac{s}{E}\right)
^{i4M\omega } 
\]

\noindent where we took the branch cut of the function $x^{i\omega }$ to run
from zero along the negative $x$ axis, wrote $x^{i\omega }=\exp \left(
i\omega \left( \ln x+i2N\pi \right) \right) $ and chose the branch $N=0$.
Thus 
\begin{equation}
i\int_{0}^{\infty }dse^{-\omega ^{\prime }s}e^{-i\omega g(is)}=i\exp \left( -%
\frac{\pi }{2}4M\omega \right) E^{-i4M\omega }\int_{0}^{\infty }dse^{-\omega
^{\prime }s}\left( s\right) ^{i4M\omega }=i\exp \left( -\frac{\pi }{2}%
4M\omega \right) E^{-i4M\omega }\frac{\Gamma \left( 1+i4M\omega \right) }{%
\left( \omega ^{\prime }\right) ^{1+i4M\omega }}  \label{co46}
\end{equation}

\noindent We substitute (\ref{co45}) and (\ref{co46}) in (\ref{co44}) and
then in (\ref{bb24}) to get

\begin{equation}
\beta _{1}(\omega ,\omega ^{\prime })=-i\frac{e^{-i\omega v_{H}}}{2\pi \sqrt{%
\omega \omega ^{\prime }}}\exp \left( -\frac{\pi }{2}4M\omega \right)
E^{-i4M\omega }\frac{\Gamma \left( 1+i4M\omega \right) }{\left( \omega
^{\prime }\right) ^{i4M\omega }}-\frac{e^{-i\omega f(0)}}{i2\pi \sqrt{\omega
\omega ^{\prime }}}  \label{co47}
\end{equation}

\noindent Collecting (\ref{bb10}) and (\ref{co47}) we get 
\begin{equation}
\beta \left( \omega ,\omega ^{\prime }\right) =-i\frac{e^{-i\omega v_{H}}}{%
2\pi \sqrt{\omega \omega ^{\prime }}}\exp \left( -\frac{\pi }{2}4M\omega
\right) E^{-i4M\omega }\frac{\Gamma \left( 1+i4M\omega \right) }{\left(
\omega ^{\prime }\right) ^{i4M\omega }}-\frac{e^{-i\omega f(0)}}{i2\pi \sqrt{%
\omega \omega ^{\prime }}}+\frac{1}{2\pi i\sqrt{\omega \omega ^{\prime }}}
\label{co48}
\end{equation}

The day is saved by (\ref{co23}) which causes an exact cancellation of the
last two terms. The first term on its own immediately leads to the black
body spectrum. Taking its modulus, squaring, and using the property 
\[
\left| \Gamma \left( 1+iy\right) \right| ^{2}=\pi y/\sinh \left( \pi
y\right) 
\]

\noindent we get 
\[
\left| \beta \left( \omega ,\omega ^{\prime }\right) \right| ^{2}=\frac{4M}{%
2\pi \omega ^{\prime }}\frac{1}{e^{8\pi \omega M}-1} 
\]

\section{Conclusion}

The objective of the paper was to prove that the Bogolubov amplitude $\beta
(\omega ,\omega ^{\prime })$ has the asymptotic form (\ref{bb2a}) and that
the radiation emitted has the spectrum of a black body. Standard quantum
mechanics dictate that we should specify the initial and final states before
calculating the transition amplitude, and to this end we considered the
photon {\it in }states before the collapse and{\it \ }the {\it out }states
after collapse has taken place (the final state does not have a $\bar{q}%
_{\omega }(u,v)$ component; cf the remark following (\ref{bb4})). We
consider the case where the gravitational field is initially weak so that we
may use the simple modes (\ref{e99}) as {\it in }modes; this is standard
practice and leads to expression (\ref{bb24}) for the amplitude.
Unfortunately the term (\ref{co80}) and the consequent (\ref{bb10}) are
missing from the standard treatments. Term (\ref{co80}) does not depend on
the kinematics of the collapse and is certainly there in order to give a
diagonal {\it S }matrix in the trivial case where collapse is never
initiated and nothing is produced. A correct treatment of the large $\omega
^{\prime }$ asymptotics of the amplitude (\ref{bb24}) yields one
contribution that leads to the black body spectrum and a second contribution
(which again is missed in the standard treatments; see Appendix A) that
precisely cancels (\ref{bb10}). However this second contribution {\it does }%
depend on the collapse and the exact cancellation takes place only in the
case of an initially weak gravitational field. In the light of the above
remarks the black body result is indeed idependent of the details of the
collapse (as often asserted) but does depend on the assumption of an
initially weak gravitational field. The question as to what would happen in
the case of an initial photon state in a gravitational background
corresponding to an advanced state of collapse is not answered either here
or in the standard treatments of Hawking radiation. The question may be of
academic interest in the context of gravity, but it may be relevant in other
cases where the analog of Hawking radiation is expected to occur.

The derivation of the black body spectrum presented in this note is based on
the calculation of the Bogolubov $\beta \left( \omega ,\omega ^{\prime
}\right) $ amplitude. As emphasized in the Introduction this quantity is by
definition time-independent, and thus the question as to where and when the
photons are produced simply does not arise. It is certainly true that were
it not for the singularity on the horizon the thermal spectrum would not
arise. There are of course arguments based on calculation of {\it local }%
field quantities to the effect that the regime in question is the important
one; or even simpler classical arguments related to the red shift during
collapse ((Weinberg 1972), p. 347). However such statements may be
misleading in connection to the quantum mechanical calculation of global (%
{\it time independent) }quantities. Similarly attempts to distinguish
between ''transient'' and ''steady state'' radiation at the level of the $%
\alpha $ and $\beta $ amplitudes are bound to fail; the emphasis in the
literature on the behaviour of the amplitude near the horizon has
unfortunately led to such statements. One of the main conclusions of this
note is that the correct derivation of the thermal result requires the
consideration of the function $f(v)$ throughout its range and not just of
its asymptotic part.

\noindent \noindent {\bf Acknowledgment}

The author is greatly indebted to Professor S\ A\ Fulling for \noindent
correspondence.

\noindent {\bf Appendix A: Comparison with previous derivations}

The black body result is often obtained via a sequence of somewhat peculiar
mathematical steps. One often starts (see e.g. (Birrell and Davies 1982) p.
108 and also references cited therein) with expression (\ref{bb24}) thus
wrongly disregarding (\ref{bb10}). Following that step one unaccountably
uses (\ref{co42}) throughout the range of integration and not just
asymptotically near the horizon where it is valid. One thus (incorrectly)
gets dropping constant prefactors 
\begin{equation}
\beta _{1}(\omega ,\omega ^{\prime })\approx \int_{0}^{v_{H}}dx\left( \frac{x%
}{E}\right) ^{i\omega 4M}e^{i\omega ^{\prime }x}  \label{a6}
\end{equation}
That (\ref{a6}) is a wrong approximation to the original integral may easily
be seen by the fact that the use of (\ref{co42}) has changed the behaviour
of the integrand at $x=v_{H}$ (the non-singular end) and an asymptotic
estimate similar to that given in section 4 would not lead to the
cancellation that took place between the last two terms in (\ref{co48}). One
then proceeds to rewrite (\ref{a6}) by rescaling $\omega ^{\prime
}x\rightarrow x$%
\begin{equation}
\beta _{1}(\omega ,\omega ^{\prime })\approx E^{-i\omega 4M}\left( \omega
^{\prime }\right) ^{-i\omega 4M-1}\int_{0}^{\omega ^{\prime
}v_{H}}dxx^{i\omega 4M}e^{ix}  \label{680a}
\end{equation}

\noindent Since one is chasing the ultraviolet divergence one simply sets $%
\omega ^{\prime }v_{H}=\infty $, changes variable $\rho =i\sigma $ and
rotates in the complex plane to get (\ref{680a}) in the form 
\begin{equation}
\beta _{1}(\omega ,\omega ^{\prime })\approx E^{-i\omega 4M}\left( \omega
^{\prime }\right) ^{-i\omega 4M-1}e^{-\frac{\pi }{2}\omega
4M}\int_{0}^{\infty }d\sigma e^{-\sigma }\sigma ^{i\omega }  \label{e18}
\end{equation}

\noindent Note that setting $\omega ^{\prime }v_{H}=\infty $ certainly does 
{\it not }amount to a systematic expansion in $\left( \omega ^{\prime
}\right) ^{-1}$. The $\sigma $ integration yields $\Gamma (1+i\omega )$ and
one thus obtains the form for the $\beta _{1}$ amplitude leading to the
black body spectrum. On the other hand the step from (\ref{680a}) to (\ref
{e18}) is again incorrect. Integral (\ref{a6}) can be performed exactly in
terms of the confluent hypergeometric function and the asymptotic estimate
for large $\omega ^{\prime }$ may be examined afterwards. Indeed let us
rescale the variable in (\ref{a6}) $x\rightarrow x/v_{H}$ and rewrite 
\begin{eqnarray}
\beta _{1}(\omega ,\omega ^{\prime }) &\approx &v_{H}^{i\omega
4M+1}E^{-i\omega 4M}\int_{0}^{1}dxe^{i\omega ^{\prime }v_{H}x}x^{i\omega 4M}=
\label{bb18} \\
&=&v_{H}^{i\omega 4M+1}E^{-i\omega 4M}\frac{1}{i\omega +1}M\left( 1+i\omega
4M,2+i\omega 4M,i\omega ^{\prime }v_{H}\right)  \nonumber
\end{eqnarray}
where $M$ is the confluent hypergeometric. We can now examine the asymptotic
limit of (\ref{bb18}) for large $\omega ^{\prime }$. The asymptotic limit of
the confluent $M(a,b,i\left| z\right| )$ for large values of $\left|
z\right| $ is given by item 13.5.1 of (Abramowitz and Stegun 1972) ($z\equiv
i\omega ^{\prime }v_{H}$). In the case $b=a+1$ some simplifications occur
and we get 
\begin{equation}
M\left( 1+i\omega ,2+i\omega ,i\left| z\right| \right) \simeq -\left(
1+i\omega \right) e^{i\left| z\right| }\frac{i}{\left| z\right| }+i\Gamma
\left( 2+i\omega \right) \frac{e^{-\frac{\pi \omega }{2}}}{\left| z\right|
^{1+i\omega }}  \label{bb20}
\end{equation}
(other terms are down by higher powers of $1/\left| z\right| $). The second
term of the above relation combined with the prefactors in (\ref{bb18}) does
feature the $\Gamma \left( 1+i\omega \right) e^{-\frac{\pi \omega }{2}}$
factor characteristic of the black body spectrum. The reason for the
discrepancy between (\ref{e18}) and (\ref{bb20}) lies in the fact that one
should first evaluate the integral in terms of the confluent and then take
the $\omega ^{\prime }\rightarrow \infty $ limit rather than take the limit
first. The rotation in the complex plane stumbles upon the Stokes phenomenon
for the confluent (different limits for $\left| z\right| \rightarrow \infty $
depending on $\arg z)$). In short the black body spectrum (\ref{e18}) is
obtained by (a) incorrectly dropping (\ref{bb10}), (b) incorrectly
approximating (\ref{bb24}) by (\ref{a6}), (c) wrongly estimating the $\omega
^{\prime }$ asymptotics of the latter.

{\bf Appendix B: Technical remarks on the collapse of a sphere of dust}

We write down the expression for the Regge-Wheeler coordinate of the stellar
radius as given by (\ref{co6}), (\ref{co11}): 
\[
R^{*}\left( \eta \right) =\frac{a}{2}\left( 1+\cos \eta \right) +2M\ln
\left[ \frac{a}{4M}\left( 1+\cos \eta \right) -1\right] 
\]

\noindent It takes a sequence of trivial trigonometric transformations to
bring this to the form 
\begin{equation}
R^{*}\left( \eta \right) =\frac{a}{2}\left( 1+\cos \eta \right) +2M\ln
\left[ \frac{\sin \frac{\eta _{0}+\eta }{2}\sin \frac{\eta _{0}-\eta }{2}}{%
\cos ^{2}\frac{\eta _{0}}{2}}\right]  \label{a1}
\end{equation}

We similarly transform the argument of the logarithm in (\ref{co8}): 
\begin{equation}
\bar{t}\left( \eta \right) =2a\sqrt{1-ka^{2}}\left[ \frac{1}{2}\left( \eta
+\sin \eta \right) +ka^{2}\eta \right] +2M\ln \left[ \frac{\sin \frac{\eta
_{0}+\eta }{2}}{\sin \frac{\eta _{0}-\eta }{2}}\right]  \label{a2}
\end{equation}

\noindent Both $R^{*}\left( \eta \right) $ and $\bar{t}\left( \eta \right) $
diverge as $\eta \rightarrow \eta _{0}$ (i.e. as the stellar radius
approaches the Schwarzschild radius), but the combination $v=\bar{t}+R^{*}$
does not: 
\begin{equation}
v=2a\sqrt{1-ka^{2}}\left[ \frac{1}{2}\left( \eta +\sin \eta \right)
+ka^{2}\eta \right] +a\cos ^{2}\frac{\eta }{2}+2M\ln \left[ \frac{\sin
^{2}\left( \frac{\eta _{0}+\eta }{2}\right) }{\cos ^{2}\frac{\eta _{0}}{2}}%
\right]  \label{a3}
\end{equation}

The quantity $C_{2}$ in (\ref{co33}) is the derivative of the above
evaluated at $\eta _{0}$: 
\begin{equation}
C_{2}=2a\sqrt{1-ka^{2}}\left( \frac{1}{2}+\frac{1}{2}\cos \eta _{0}\right) -%
\frac{a}{2}\sin \eta _{0}+2M\cot \eta _{0}  \label{a4}
\end{equation}

The quantity $C_{1}$ in (\ref{co33}) is trivially obtained by
differentiating (\ref{co7}): 
\begin{equation}
C_{1}=\frac{a}{2}\sin \eta _{0}  \label{a5}
\end{equation}

To calculate $A$ in (\ref{co14}) we rewrite the latter in the form 
\begin{equation}
\frac{\bar{R}-2M}{2M}\simeq Ae^{-\bar{t}/2M}  \label{a7}
\end{equation}

\noindent We take logarithms of both sides of (\ref{a7}) to rewrite it in
the form 
\begin{equation}
\ln \left( \frac{\bar{R}-2M}{2M}\right) \simeq \ln A-\frac{\bar{t}}{2M}
\label{a9}
\end{equation}
We now use (\ref{co5}), (\ref{co9}), (\ref{a9}) and a sequence of
trigonometric identities to obtain 
\begin{equation}
\ln \left[ \frac{2}{ka^{2}}\sin \eta _{0}\right] =\ln A-\frac{a}{M}\sqrt{%
1-ka^{2}}\left[ \frac{1}{2}\left( \eta _{0}+\sin \eta _{0}\right)
+ka^{2}\eta _{0}\right]  \label{a8}
\end{equation}

\noindent thus determining $A$. The crucial step in the calculation of $A$
lies in the fact that although both sides of (\ref{a9}) diverge in the limit 
$\eta \rightarrow \eta _{0}$, there is precise cancellation of a term 
\[
\ln \left( \sin \frac{\eta _{0}-\eta }{2}\right) 
\]

\noindent on each side thus leading to the finite result (\ref{a8}).

For a given value of $\eta _{0}$ the corresponding value $t(\eta _{0})$
given by (\ref{co7}). We take advantage of the fact that in the comoving
frame the endpoints B and A of the ray BCA lie at $r=a$. Thus the unknown is
the time $t_{H}$ in the comoving frame where ray $l1$ hits the star's
surface. The latter may be determined via the equation of a null geodesic
inside the star obtained through (\ref{c1}). The path of a light ray BCA
propagating inside the star is given by 
\begin{equation}
\int_{t_{H}}^{t_{0}}\frac{dt}{R(t)}=2\int_{0}^{a}\frac{dr}{\sqrt{1-kr^{2}}}
\label{co15}
\end{equation}

\noindent To evaluate the left hand side of (\ref{co15}) we change variable
according to (\ref{co7}) 
\[
\frac{d\tau }{d\eta }=\frac{1}{2\sqrt{k}}\left( 1+\cos \eta \right) 
\]
Using (\ref{co6}) for $R(t)$ the left hand side of (\ref{co15}) immediately
yields 
\begin{equation}
\frac{1}{4\sqrt{k}}\left( \eta _{0}-\eta _{H}\right)  \label{co30}
\end{equation}
The right hand side of (\ref{co15}) yields 
\[
\frac{2}{\sqrt{k}}\arcsin \left( \sqrt{k}a\right) 
\]
which in the limit (\ref{co31}) reduces to $2a$. Combining with (\ref{co30})
we get 
\begin{equation}
\eta _{0}-\eta _{H}=8\sqrt{k}a  \label{co56}
\end{equation}
In the limit (\ref{co31}) we get that $v_{H}$ lies very close to $v_{0}$.

\end{document}